\documentclass[pre]{revtex4-2} 
\usepackage{graphicx}
\usepackage{dcolumn}
\usepackage{bm}
\usepackage{xcolor}
\usepackage{comment}
\usepackage[mathlines]{lineno}
\usepackage{hyperref}
\hypersetup{
    colorlinks,
    citecolor=black,
    filecolor=black,
    linkcolor=blue,
    urlcolor=black
}

\begin{document}
\title[]{Fluctuations of local plastic strain in granular media}
\author{I. Awada, M. Bornert, V. Langlois, J. Léopoldès}%
 \email{julien.leopoldes@univ-eiffel.fr}
\affiliation{ 
Université Gustave Eiffel, ENPC, Institut Polytechnique de Paris, CNRS, Navier, 77454 Marne-la-Vallée, France
}%
\date{\today}
            
\begin{abstract}
We experimentally study the heterogeneity of strain in a granular medium subjected to oscillatory shear in a rotating drum. Two complementary methods are used. The first method relies on optical imaging and grain tracking, allowing us to compute some components of the strain tensor and their variance. The second method, Diffusive Acoustic Wave Spectroscopy (DAWS), provides the quadratic strain within the bulk. Our results show that strain is spatially heterogeneous, with fluctuations about ten times larger than the mean, primarily dominated by variability at the grain scale. We then analyze in detail the strain fluctuations occurring during the forward and backward branches of the shear \textcolor{black}{stress cycles, along with the intra-cycle plastic strain resulting from of each cycle}. Both methods reveal that each shear cycle consists of two consecutive diffusive-like branches, and that the resulting plastic strain fluctuations scales with the mean plastic shear strain. \textcolor{black}{We propose that plastic strain fluctuations result from irreversible strain heterogeneity that increases with applied shear---reflected in forward--backward strain anticorrelations---but is constrained by load-controlled induced memory.}

\end{abstract}
\maketitle
\section{Introduction}
Solids respond elastically to shear at low strain, then progressively yield and eventually fail. In heterogeneous and discrete solid systems composed of particles, such as foams, emulsions, colloidal glasses, or granular media, macroscopic plasticity develops through complex mechanisms at the scale of one to ten particle diameters. For example, rearrangements in foams and emulsions are known as T1 events (neighbor switching between four particles)~\cite{argon2,chen}. In colloidal glasses, Eshelby-like inclusions are responsible for plastic strain~\cite{spaepenW}. In granular media, strain is highly non-affine, as evidenced by the emergence of vortex-like structures~\cite{combe2015,radjai2002} and the presence of defects similar to T1 events~\cite{cao2018}. Still, how the intensity of plastic strain fluctuations depends on system packing density and applied strain/stress remains poorly understood in three-dimensional systems.

In this context, it is essential to develop experimental methods capable of resolving particle rearrangements in granular media. Bulk particle rearrangements can be investigated using scanning techniques such as laser sheet imaging~\cite{pouliquen2003}, confocal microscopy~\cite{spaepenW}, or X-ray tomography~\cite{hurley}. Additionally, diffusing wave spectroscopy (DWS) is a well-established method for studying divided matter, providing estimates of the random relative displacement between particles such as drops, colloids, or bubbles~\cite{hohler1997,munch97,petekidis2002}. In granular materials, diffusive light scattering has been extensively used to probe interactions between strain heterogeneities at the scale of about $10$ grain diameters~\cite{crassous2,crassous1} and to analyze velocity fluctuations in the liquid state~\cite{durian_97}. Since the wavelength of ultrasound can be easily adjusted to match the size of a grain, Diffusive Acoustic Wave Spectroscopy (DAWS) enables the measurement of kinematics at the grain scale~\cite{weitzpage}, revealing the transition from reversible to irreversible local grain motions at the macroscopic flow threshold~\cite{leosoft2020}. However, applying DAWS to sheared granular media has not yet enabled a quantitative analysis of strain fluctuations.  

This paper has two aims: to introduce a method for monitoring grain-scale strain fluctuations in sheared granular media and to provide new insights into how the magnitude of local strain heterogeneity depends on the applied macroscopic shear.
\section{Experiments}
\textcolor{black}{The system consists of a dense granular material immersed in water and subjected to oscillatory rotational motion. This setup is relevant to soil creep~\cite{culling}, with its associated natural hazards, as well as to diffusion-induced granular mixing~\cite{choi}.} The granular material is placed inside a rotating drum that oscillates slowly at angles lower than the angle of repose. Given the quasi-static loading, the liquid water does not influence grain rearrangement.  

Two series of experiments were conducted using the same preparation protocol but different measurement techniquesmethods. The first method involves analyzing the two-dimensional kinematics by tracking the positions of grains located at the front face of the drum. The second method, DAWS, is a three-dimensional probe of the granular medium and relies on the correlation of ultrasound signals propagating through the water pores.  

\subsection{Preparation protocol of the initial state and oscillations}
\label{prep}
\begin{figure}
    \centering
    \includegraphics[scale=0.7]{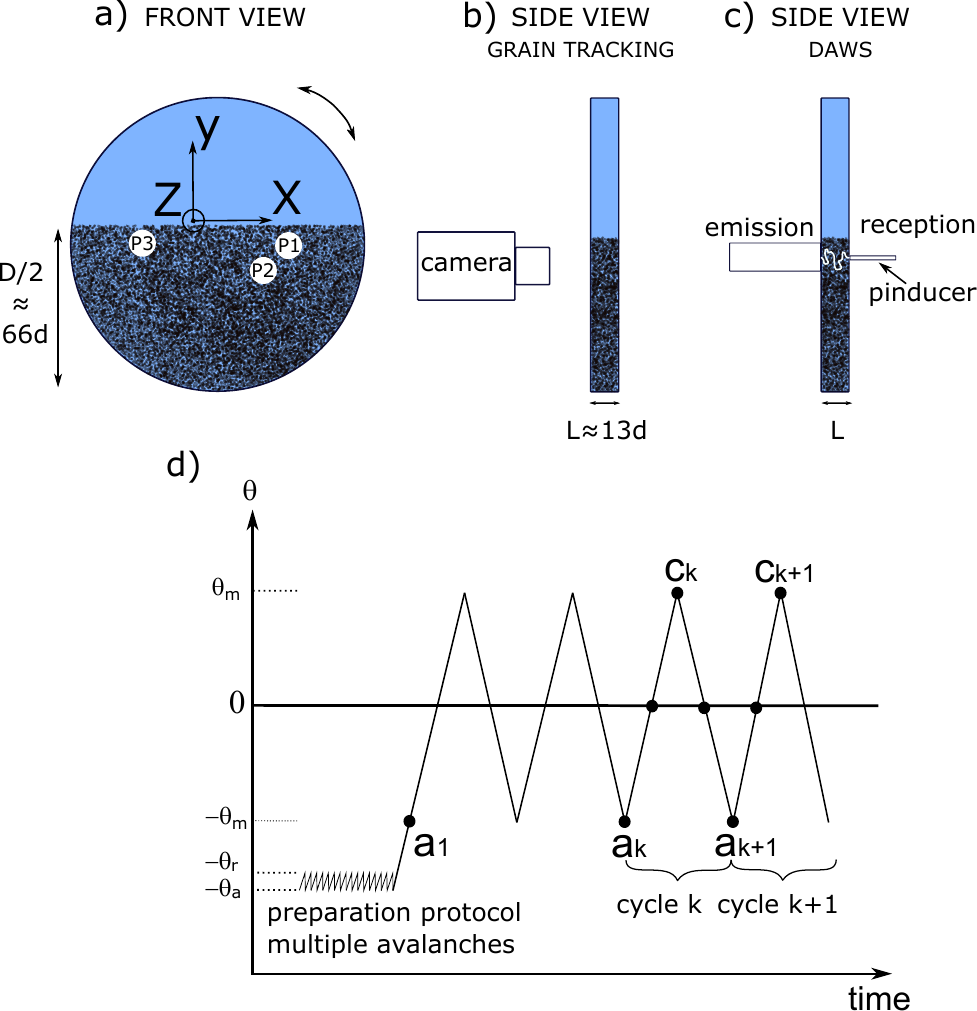}
    \caption{\textcolor{black}{Sketch of the experiment. The rotating drum (\textcolor{black}{radius $D/2 = 100$ mm and thickness $L = 20$ mm}) is filled with grains and water. (a) The front view shows the grains and the three zones of interest analyzed. (b) The side view for the grain tracking method shows the camera (Basler acA1920), which rotates with the drum. Only the grains at the front face are imaged. (c) The side view for the DAWS setup illustrates the ultrasound emission and reception transducers. The thick white line represents an arbitrary diffusive path. (d) For both the grain tracking method and \textcolor{black}{DAWS}, the preparation protocol begins with the continuous rotation of the drum, producing several consecutive avalanches. Then, the free surface is brought to an angle $-\theta_m$, lower than the angle of repose, and subjected to 50 oscillations between $-\theta_m$ and $+\theta_m$. Black points indicate the angles at which images or ultrasonic signals are recorded. The notations $a_k$ and $c_k$ are introduced later in the text to define the pairs of images (two granular configurations) between which the strain is analyzed.}}
    \label{fig:grains-pictures}
    
\end{figure}

A sketch of the rotating drum is provided in Figure~\ref{fig:grains-pictures} a), b), \textcolor{black}{and c)}. After introducing the grains (ceramic, diameter $d=1.50\pm0.08$ mm), the drum is mounted on a rotating stage. It is then filled with water, ensuring the removal of any air bubbles. Each experiment is preceded by a rapid, continuous anticlockwise rotation of the drum by at least $180^\circ$. The preparation protocol then consists of a slower clockwise rotation, inducing approximately twenty consecutive avalanches (Figure~\ref{fig:grains-pictures} d)). This quasi-static rotation allows us to measure two characteristic angles, defined as the inclination $\theta$ of the granular free surface relative to the horizontal plane.  
The avalanche angle, $\theta_a=25.5^\circ$, represents the maximum inclination before an avalanche occurs, while $\theta_r=22.8^\circ$ is the inclination at which the granular medium stops flowing~\cite{repose}. Consequently, the preparation protocol ensures that the free surface of the granular medium is initially inclined at $\theta \approx \theta_r$. 

At this stage, the drum is rotated so that the free surface is set to an inclination $\theta_m<\theta_r$ (Fig.~\ref{fig:grains-pictures}d)). The system is then subjected to 50 oscillations at $\theta=\pm \theta_m$, indexed by $1\leq k\leq 50$. \textcolor{black}{The granular medium is therefore studied in the plastic "solid" regime. As will be shown later, this corresponds to a typical shear strain of $\epsilon\sim 10^{-3} \rightarrow 10^{-2}$ smaller than the yield strain but much larger than the strain characteristic of the elastic regime where $\epsilon\sim 10^{-6}$.} \textcolor{black}{To mitigate inertial effects, the rotation between positions labeled $a_k$ and $a_{k+1}$ (Fig.~\ref{fig:grains-pictures}d)) is performed in four phases combining constant---and deliberately small---acceleration and deceleration stages: (i) acceleration at $0.2^\circ/\mathrm{s}^2$, (ii) stationary rotation phase at a constant angular velocity of $0.5^\circ/\mathrm{s}$, and (iii) deceleration phase at $0.2^\circ/\mathrm{s}^2$. Furthermore, after reaching the target angular position (black dots Fig.~\ref{fig:grains-pictures}d)), a waiting period of $30~\mathrm{s}$ is imposed to ensure all grain motion fully relaxes before acquisition. Ten consecutive images are taken at each position to allow for averaging and reduce residual noise.} We present results for four different values of $\theta_m$: $21.8^\circ$, $19.8^\circ$, $16.8^\circ$, and $10.8^\circ$. \textcolor{black}{For ensemble averaging, the entire preparation protocol and the subsequent fifty shear cycles are repeated 10 times for each $\theta_m$.}

\subsection{Image analysis}
The granular medium is characterized by capturing four images per cycle. An example of such an image is shown in Figure~\ref{fig:image}~a). \textcolor{black}{The camera is rigidly mounted on the frame of the rotating stage, so that images are recorded in the reference frame of the camera.} We present data for three different areas of interest, \textcolor{black}{fixed on the side of the drum}. Two of these areas are near the free surface (P$1$ and P$3$), while the third one is located deeper within the packing (P$2$), as shown in Figure~\ref{fig:image}a).  
\subsubsection{Grain detection and tracking}
\begin{figure}
    \centering
    \includegraphics[scale=0.11]{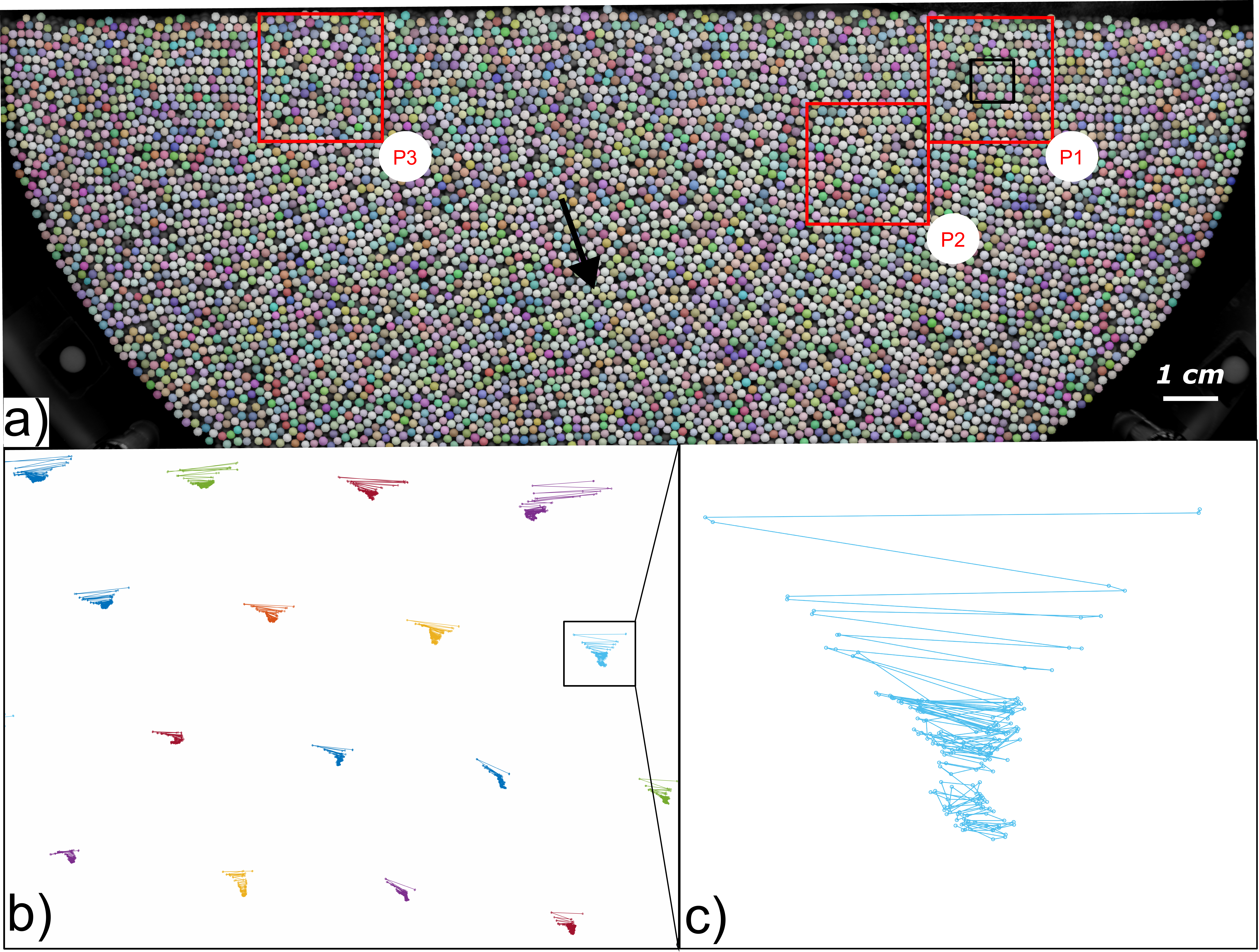}
    \caption{a) Typical image taken at the front face of the drum, illustrating examples of grain detection the end of the preparation protocol for $\theta_m=19.8^\circ$. \textcolor{black}{The black arrow represents the direction of gravity $\vec g$ at that angle.} Each detected grain is arbitrarily colored. The positions "$P1$", "$P2$", and "$P3$", outlined by red squares, correspond to the zones of interest where deformation is computed and compared with the decorrelation of acoustic signals. b) Reconstructed trajectories after tracking for 50 cycles at $P1$ and oscillation angle $\theta_m=19.8^\circ$. The field of view corresponds to the black square in a. c) Zoom on a selected trajectory from b). \textcolor{black}{The first point $a_3$ is located on the top-right part of the image.}
    }
    \label{fig:image}
\end{figure}

For each image, we apply a deep learning method to detect the grains and segment the image~\cite{stardist}. To train the model, we performed a meticulous manual annotation of grain contours on a series of images. From the segmentation results, the center of each grain was determined as the weighted average position of the pixels in the detected region, with weights given by the gray levels. Subsequently, a tracking algorithm~\cite{track} was used to reconstruct grain trajectories during the drum oscillations. A selection of the reconstructed trajectories is shown in Figure~\ref{fig:image}~b) and~c). \textcolor{black}{The grains oscillate as they gradually sink under the effect of gravity.}  

\textcolor{black}{During the initial oscillations, grain motion can be significant ($>d/2$), leading to inaccurate tracking. To ensure robust analysis, we retain only images where grain tracking can be visually validated. The first cycle in which grains are accurately tracked is $k_r=4,3,2,1$ for $\theta_m=21.8^\circ, 19.8^\circ, 16.8^\circ$, and $10.8^\circ$, respectively. This first cycle is marked with a star in Figure~\ref{fig:freeS}a).}  

\subsubsection{Strain analysis and averaging procedure}
The set of grain coordinates is used to construct a Delaunay triangulation, where the triangle vertices correspond to the centers of mass of the grains. The various components of the strain (symmetric part of the strain tensor) are then computed at the grain scale following Bagi~\cite{bagi96}. Throughout this paper, these triangles will be referred to as "Bagi cells" (strain noise level $10^{-4}$). An example of such Bagi cells, \textcolor{black}{color-coded according to the strain magnitude, is shown} in Figure~\ref{fig:fluctuations}a). We have verified (not shown) that the average volumetric strain, weighted by area and computed over the entire drum, matches the volumetric strain derived from the displacement of the granular free surface.

\textcolor{black}{The 2D strain in Bagi cells is averaged over bins (zones $P1$, $P2$, $P3$) of size $13~d\times13~d$, as shown by the red squares in Figure~\ref{fig:image}a). This spatial average is denoted by $\overline{\cdot}$. The size of these zones is chosen to match the surface area illuminated by ultrasound waves in the DAWS method (see Appendix~\ref{sec:daws}). Other averaging zone sizes will be used occasionally and will be explicitly mentioned.} 

\textcolor{black}{Because the Bagi cell strain measurements are obtained from experiments different from those using DAWS (as the acoustic sensors obstruct the field of view), comparing the results from both methods also requires ensemble averaging.} This ensemble averaging is performed over 10 experiments, and the resulting strains are denoted as $\langle \overline{\cdot} \rangle$. Additionally, some correlations between individual Bagi cells will be computed without averaging.

\textcolor{black}{The optical imaging method allows us to compute the strain components $\langle \overline{\epsilon_{xx}} \rangle$, $\langle \overline{\epsilon_{yy}} \rangle$, $\langle \overline{\epsilon_{xy}} \rangle$, and $\langle \overline{\epsilon_{v}} \rangle$ \textcolor{black}{(volumetric)}, while $\langle \overline{\epsilon_{zz}} \rangle=0$ due to the presence of solid walls. In addition to being limited to two dimensions, optical imaging captures the kinematics of grains at the front face of the drum, which may behave differently from those within the bulk. For instance, it is well known that in flowing granular media, solid walls induce layering, density oscillations, and inhomogeneous granular temperature~\cite{savage93}.}

\textcolor{black}{The strain is computed from a pair of granular configurations $p_1$ and $p_2$ selected from the set $\left\{a_k,c_k\right\}$ with $k_r\leq k$. This pair is denoted as $\left[p_1,p_2\right]$, where $p_1$ serves as the reference configuration.}

\subsubsection{Free surface}
A useful analysis is the detection of the free surface to estimate the variation in the global solid fraction throughout the experiments. 
We first divide the image into 30 vertical stripes of equal width (4d). Then, we compute the profile of intensity for each stripe and apply a detection threshold to find the position of the free surface $y_s$ for each stripe. \textcolor{black}{We calculate its average position throughout the width of the drum $\overline{y_s}$ followed by ensemble averaging. The vertical displacement of the free surface from its initial position that was reached at the end of the protocol of preparation is defined as $\langle \overline{-s_y} \rangle =\langle \overline{y_s(k)} \rangle-\langle \overline{y_s(0)} \rangle$.} \textcolor{black}{Unlike grain tracking, free surface detection can be performed as early as the first oscillation cycle. For comparison, the first cycles from which grain tracking could be performed are annotated by stars in Figure~\ref{fig:freeS} b).}

\subsection{Diffusing acoustic wave spectroscopy}
\label{subsection:us}
It is also \textcolor{black}{possible} to probe the granular medium in the bulk, which is achieved using DAWS. When the wavelength of a propagating wave is on the order of the particle size, multiple scattering occurs due to the numerous scatterers encountered along the wave’s path. DAWS relies on analyzing the variations in the phase variance of the propagating signal along the acoustic path of $n$ scatterers (grains), caused by their relative motion~\cite{weitzpage,sheng}.  

However, in sheared granular media, interpreting the decorrelation of multiply scattered waves is complex due to strain heterogeneities across multiple length scales (grain scale, shear bands). Additionally, ubiquitous phenomena such as compaction and plasticity make ensemble averaging challenging. Below, we propose a protocol to extract meaningful kinematic information from DAWS in dense granular media, allowing for comparison with the previous method based on direct grain tracking.

The scattered waves produce signals, $\Psi_i$ and $\Psi_j$, measured for two slightly different granular configurations, which are compared by computing a correlation function $g_1$:  
\begin{equation}
    g_1=\frac{\int \Psi_i\Psi_jdt}{(\int \Psi_i^2dt\int\Psi_j^2dt)^{1/2}}
    \label{eq:gamma}
\end{equation}
$g_1$ represents the spatial average of the many scattering events occurring within the volume probed by the wave. When ultrasound is used to monitor structural changes in the granular packing under oscillations, the decorrelation of the signals reflects relative grain displacements, or strains, at the length scale of the mean free path $l \sim \lambda \sim d$~\cite{weitzpage}, \textcolor{black}{see Appendix~\ref{sec:daws}}.  

The works of Bicout et al.~\cite{bicout1994} and Cowan et al.~\cite{weitzpage} establish the relationship between \textcolor{black}{the ensemble-averaged} $g_1$ and the phase shift of a single scattering event, $\Delta \Phi$, induced by local strains:

\begin{equation}
g_1=\cos\left(n\langle \Delta \Phi \rangle\right) \times \exp\left(-\frac{n}{2}V\left(\Delta \Phi\right)\right).
\label{bicout:eq}
\end{equation}
where $\langle\Delta \Phi \rangle$ is the average phase shift, $V\left(\Delta \Phi\right)$ the variance and $n$ the number of scattering events along the acoustic path. 
In addition, the authors show that :
\begin{eqnarray}
	\langle \Delta \Phi \rangle & = &\frac{1}{3}ql\langle \epsilon_v \rangle,\label{bicout:eq2a}\\
	V\left(\Delta \Phi\right) & = &\frac{2}{15}\left(ql\right)^2\left(\langle\epsilon_v^2\rangle + 2\times \sum_{i,j} \langle \epsilon_{i,j}^2 \rangle \right),\label{bicout:eq2b}
\end{eqnarray}
where $q=2\pi/\lambda$ and $l$ is the mean free path. Hence, with $\langle \epsilon_v \rangle$ and $g_1$ we can determine the phase variance $V(\Delta \Phi)$ using Eqs.~\ref{bicout:eq} and \ref{bicout:eq2a}. This, in turn, allows us to compute the sum of quadratic terms on the right-hand side of Eq.~\ref{bicout:eq2b}, denoted as $\langle \overline{\epsilon^2} \rangle_{u}$.

In a specific set of experiments that solely involve acoustic measurements, we use the rotating drum equipped with a pair of ultrasound emitter-receiver transducers, as illustrated in Figure~\ref{fig:grains-pictures} c). In this configuration, grain tracking is not possible because the transducers obstruct the field of view. Briefly, a pulse is emitted by a 1 MHz immersion transducer (diameter $d_t = 13$~mm, Panametrics) and received by a small pinducer with a diameter of $d_p = 1.5$~mm. Given the corresponding wavelength $\lambda = 1.5$~mm, which is approximately equal to $d_p$, the waves undergo multiple scattering~\cite{weitzpage,sheng}. In the present study, DAWS probes the medium at the length scale of the grains, sampling a volume roughly corresponding to $d_t^2 \times L \sim 10^3 d^3$. More details about DAWS are provided in Appendix~\ref{sec:daws}.

\section{Results}
\subsection{Compaction}
\label{sub:compactmacro}
\begin{figure}[!]
    \centering
    \includegraphics[scale=0.5]{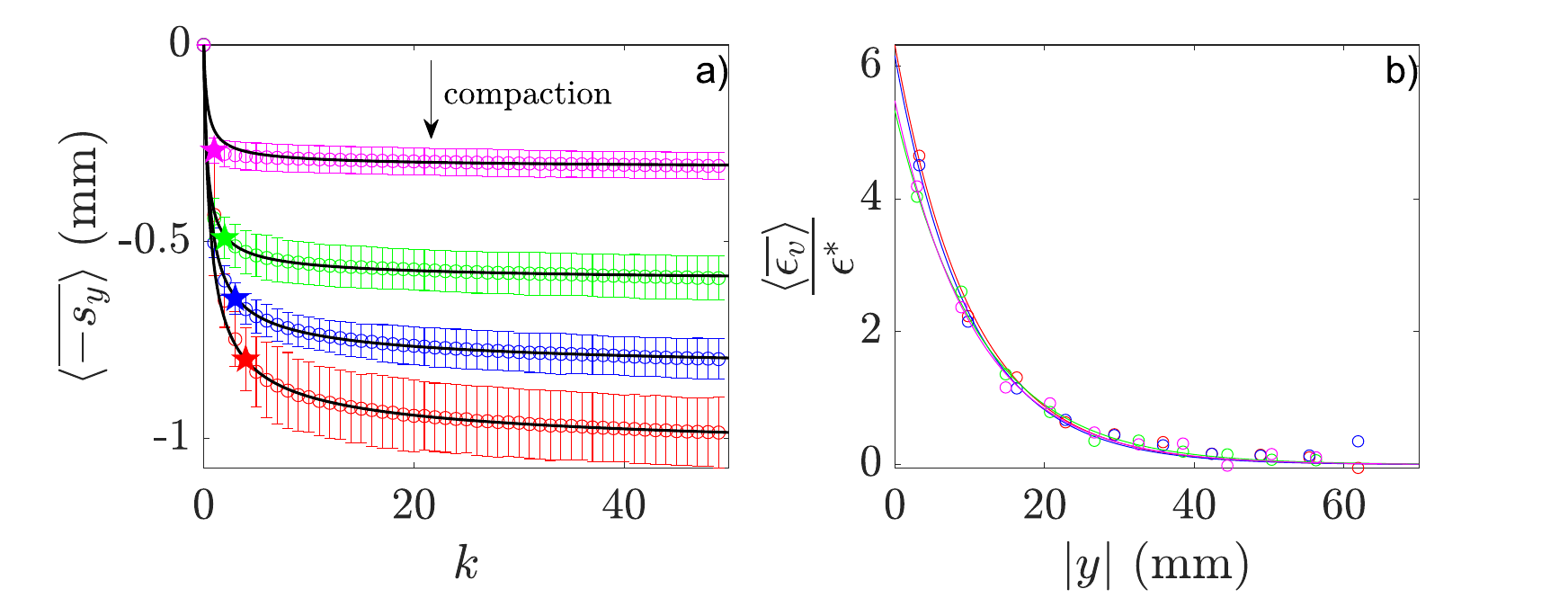}
   \caption{
   	a) Variation of the free surface position as a function of the number of oscillations. The curves, from lowest to highest, correspond to $\theta_m = 21.8^\circ$, $19.8^\circ$, $16.8^\circ$, and $10.8^\circ$. The black lines represent the fit from the model proposed by~\cite{nagel2}. The fitted parameters are provided in the appendix (see Fig.~\ref{fig:sup:fitmodelnagel}). A star denotes the first images where grain tracking could be performed. 
   	b) Profile of the volumetric strain [$a_{k_{r}}, a_{50}$] along the vertical axis $\langle \overline {\epsilon_v(\theta_m)} \rangle$, normalized by the mean value $\epsilon^*$ over the region $0 < |y| < 60~\text{mm}$. The colors red, blue, green, and pink correspond to $\theta_m = 21.8^\circ$, $19.8^\circ$, $16.8^\circ$, and $10.8^\circ$, respectively. Experimental data are represented by circles, while the thin lines represent the exponential fit (as described in the text).
   }
       \label{fig:freeS}
\end{figure}
As the number of oscillations increases, the granular material undergoes compaction, as evidenced by the variation of $\langle \overline{-s_y} \rangle$ shown in Figure~\ref{fig:freeS} a). Higher oscillation angles result in denser packings for a fixed number of oscillations. Additionally, compaction occurs more rapidly at larger oscillation angles. The evolution of packing fraction during compaction is fitted to the standard logarithmic law, as proposed by~\cite{nagel2}, and is presented in Figure~\ref{fig:freeS} a). The equation used for the fitting and the corresponding parameters are provided in Appendix~\ref{supp:paramFit}, Figure~\ref{fig:sup:fitmodelnagel}. \textcolor{black}{It can be observed that the quasi-static oscillations employed in this study produce an effect similar to vertical mechanical taps~\cite{nagel2} in terms of compacting the granular material. However, this analogy should be interpreted with caution, as the underlying mechanisms may differ: while classical tapping induces compaction through inertial driving, the present oscillations primarily promote compaction via quasi-static shearing.}

Furthermore, as shown by~\cite{nagel2}, the compaction is spatially heterogeneous, being more pronounced near the free surface than in the bulk of the granular medium. To capture this spatial variation with sufficient resolution, we apply a specific binning technique along the vertical direction by dividing the drum images into horizontal stripes of thickness $4d$, and compute the strain for each stripe. The profiles of volumetric strain between [$a_{k_{r}},a_{50}$] are shown in Figure~\ref{fig:freeS} b) as a function of depth, along with an exponential fit of the form $\langle \overline {\epsilon_v} \rangle \propto \exp \left( -\frac{y}{y^*} \right)$. A similar profile is observed for each oscillation angle $\theta_m$, with the characteristic cutoff length $y^* = 11.4 \pm 3$~mm.
\subsection{Macroscopic strain cycles and local strain fluctuations}
\subsubsection{$\theta\text{-}\epsilon$ cycles}
\label{eq:compac}
\begin{figure} 
    \centering
   \includegraphics[scale=0.6]{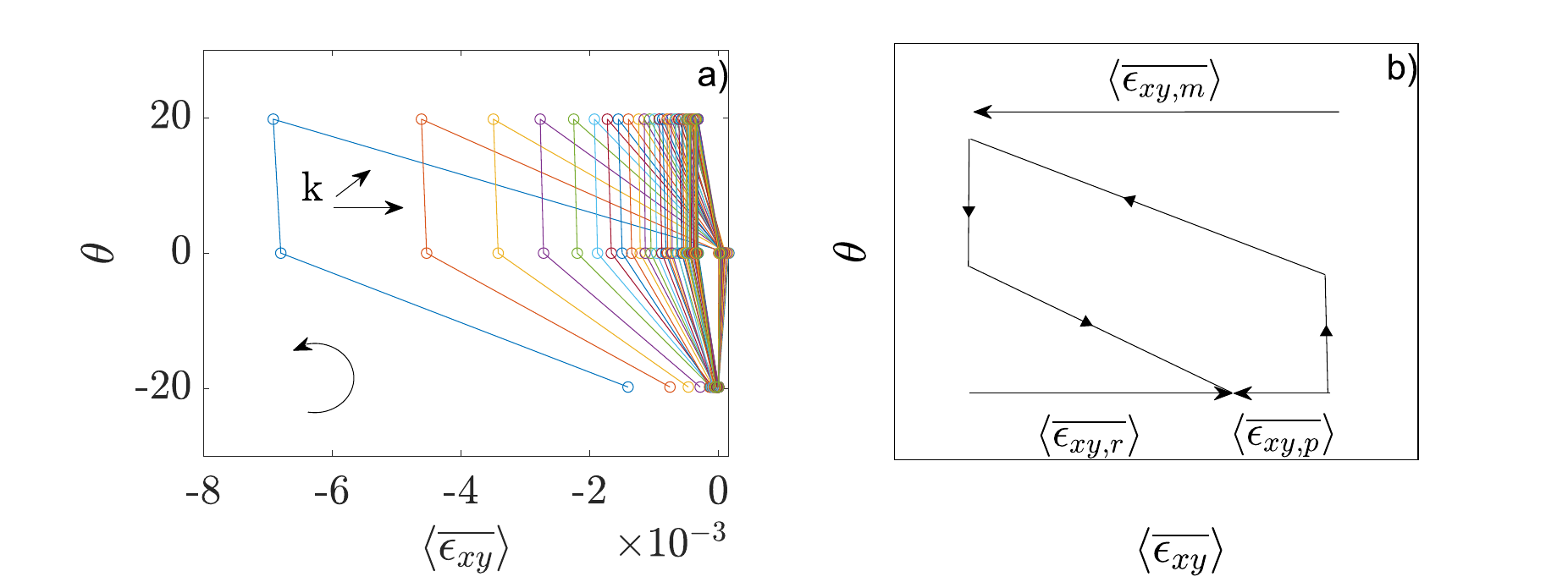}
   \caption{\textcolor{black}{a) Superposition of shear cycles at $\theta_m=19.8^\circ$ (Position $P1$, $k_r=3$). Cycles are in the anticlockwise direction, as indicated by the arrow. The first cycle (blue line) has the largest amplitude of deformation, and the following ones shift to the right. b) Sketch of a typical cycle with annotation of some intra-cycle strains as considered in the analysis (see text).}}
    \label{fig:cycle}
\end{figure}

In Figure~\ref{fig:cycle}a), we plot $\theta\text{-}\epsilon$ cycles, focusing on the shear strain $\langle\overline{\epsilon_{xy}}\rangle$ at position $P1$ for $\theta_m = 19.8^\circ$. As the granular material compacts, the maximum intra-cycle shear strain \textcolor{black}{$\langle \overline{\epsilon_{xy,m}} \rangle$}, defined as the strain between $[a_k, c_k]$ and illustrated in Figure~\ref{fig:cycle}b), decreases with the number of oscillations. 

For the free-surface configuration used in the present study, the rotation of the system by an angle $\theta_m$ is equivalent to the application of a shear stress of the order $\sigma_m \sim \Delta \rho \phi h \sin \theta_m$, where $\Delta \rho = \rho_g - \rho_w$ (with $\rho_g$ and $\rho_w$ being the densities of the grains and water, respectively), $\phi = \phi(\theta_m, k)$ is the packing fraction, and $h = 13d$ is the size of the zones. According to the preparation protocol (Figure~\ref{fig:grains-pictures} a)), all samples exhibit the same global initial packing fraction $\phi_0 \approx 0.61$, which is obtained from the position of the free surface averaged over all experiments.

To estimate the packing fraction $\phi$ of the material located in the zones $P1$ and $P3$, we use the relation $\phi \approx \phi_0 \left(1 + \langle \overline{s_y} \rangle / h\right)$, where $\langle \overline{s_y} \rangle$ is the position of the free surface, as shown in Figure~\ref{fig:freeS} a). The difference in packing fraction between two systems sheared at different $\theta_m$ is given by $\Delta \phi / \phi_0 \approx \langle \overline{s_y} \rangle / h \ll \Delta \theta_m / \theta_m$. Moreover, at a fixed $\theta_m$, the relative variation of packing fraction during compaction is small, on the order of $\sim 10^{-2}$. Therefore, $\theta_m$ determines the constant maximum stress applied at each cycle, independent of $\phi$. However, due to the repeated oscillations, $\theta_m$ do modify the packing fraction by also controlling the intensity of "mechanical taps", as discussed above. Therefore, the constitutive relation describing the cycles can be expressed as $\langle \overline{\epsilon_{xy,m}} \rangle(\theta_m,\phi(\theta_m,k))$, increasing function of $\theta_m$ at fixed $\phi$ and decreasing function of $\phi$. In turn, $\phi$ is an increasing function of both $k$ and $\theta_m$.

Additionally, between $[a_k, a_{k+1}]$, each cycle is characterized by finite plastic deformation \textcolor{black}{$\langle \overline{\epsilon_{xy,p}} \rangle$} (see also Figure~\ref{fig:cycle}b). After a large number of oscillations, the cycles stabilize and the plastic strain disappears. This stabilization, along with permanent cumulative deformation, is referred to as "accommodation" in soil mechanics~\cite{hicher}.
\subsubsection{\textcolor{black}{Strain heterogeneities.}}
\begin{figure} 
    \centering
    \includegraphics[width=1\textwidth]{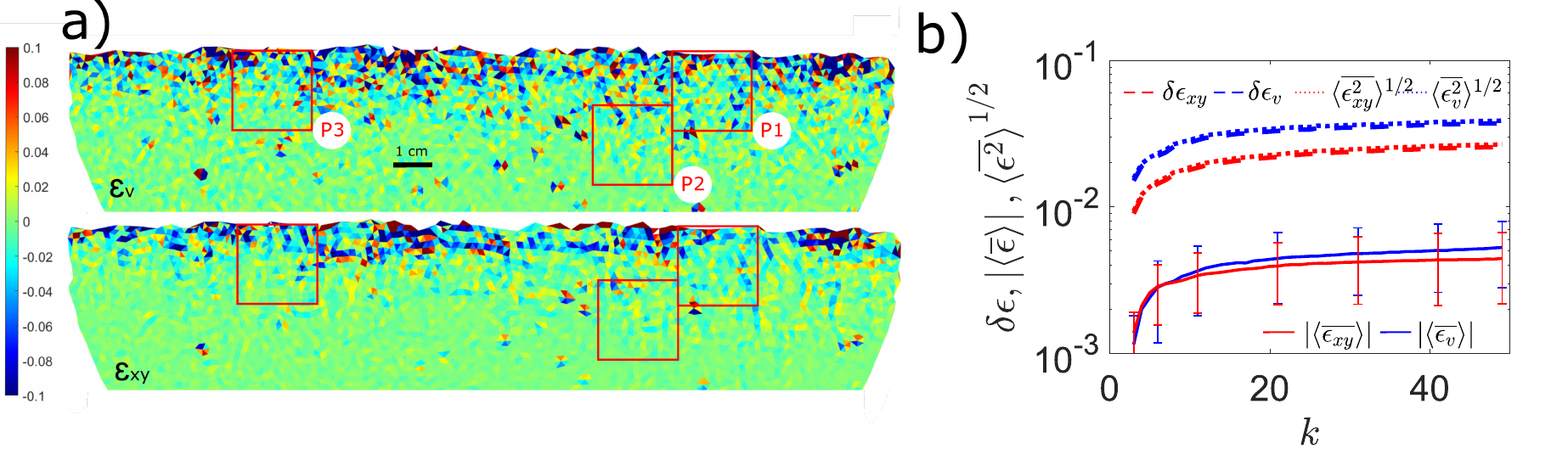}
    \caption{a) Maps of Bagi cells for \textcolor{black}{cumulative} plastic volumetric strain (upper panel) and plastic shear strain (lower panel) for $\theta_m=19.8^\circ$, pair $[a_{k_r},a_{50}]$. The upper portion of each panel corresponds to the free surface. b) Plot of \textcolor{black}{cumulative} $|\langle \epsilon_{v,p} \rangle|$, $|\langle \epsilon_{xy,p} \rangle|$, $\langle \epsilon^2_{v,p} \rangle^{1/2}$, $\langle \epsilon^2_{xy,p} \rangle^{1/2}$, and $\delta \epsilon_{v,p}$, $\delta \epsilon_{xy,p}$ as a function of the number of oscillations $k$, at position $P1$ for $\theta_m = 19.8^\circ$. Data are presented for pairs $[a_{k_r},a_{k}]$. The ordinate is shown on a logarithmic scale for convenient visualization of the data. The error bars on the $|\langle \epsilon_{v} \rangle|$ and $|\langle \epsilon_{xy} \rangle|$ curves represent the standard deviation arising from the variability in the mean deformation between different experiments (10 in total).}
    \label{fig:fluctuations}
\end{figure}

An example of a deformation map for the interval $[a_{k_r}, a_{50}]$ is shown in Figure~\ref{fig:fluctuations} a), where both volumetric and shear \textcolor{black}{cumulative plastic} strains are presented. The strain distribution is spatially heterogeneous across two length scales. At the grain scale, adjacent Bagi cells can exhibit strain differences by several orders of magnitude, with strains that can even have opposite signs. At the larger scale represented by the characteristic cutoff length $y^*$ in Figure~\ref{fig:freeS} b), the strain map shows that that strain is more pronounced near the free surface.

In Figure~\ref{fig:fluctuations} b), we present the \textcolor{black}{cumulative} strain and its root-mean-square fluctuation $\delta \epsilon_p = \left( \langle \overline{\epsilon_p^2} \rangle - \langle \overline{\epsilon_p} \rangle^2 \right)^{1/2}$ for the interval $[a_{k_r}, a_{k}]$, which corresponds to the total plastic deformation accumulated during a given cycle. For both the shear and volumetric components, the root-mean-square fluctuations are approximately \textcolor{black}{five to} ten times greater than the mean, with $\delta \epsilon_p \approx \langle \overline{\epsilon_p^2} \rangle^{1/2} \gg |\langle \overline{\epsilon_p} \rangle|$. This is not attributable to the heterogeneity of the mean strain, as an exponential profile leads to a variability of the same order of magnitude as the mean. Furthermore, the fluctuations are not influenced by the reproducibility of $\langle \overline{\epsilon_p} \rangle$ across different experiments, as shown by the error bars in Figure~\ref{fig:fluctuations} b), which are much smaller than the fluctuations themselves. Therefore, the fluctuations are primarily governed by the strain heterogeneity at the grain scale, caused by the erratic motions of the grains. These motions are the dominant source of decorrelation in the scattered waves, which is quantified by $g_1$. From this point onward, we will refer to $\delta \epsilon \approx \langle \overline{\epsilon^2} \rangle^{1/2}$ as fluctuations at the grain scale.

\begin{figure}[h]
    \centering
    \includegraphics[width=1\textwidth]{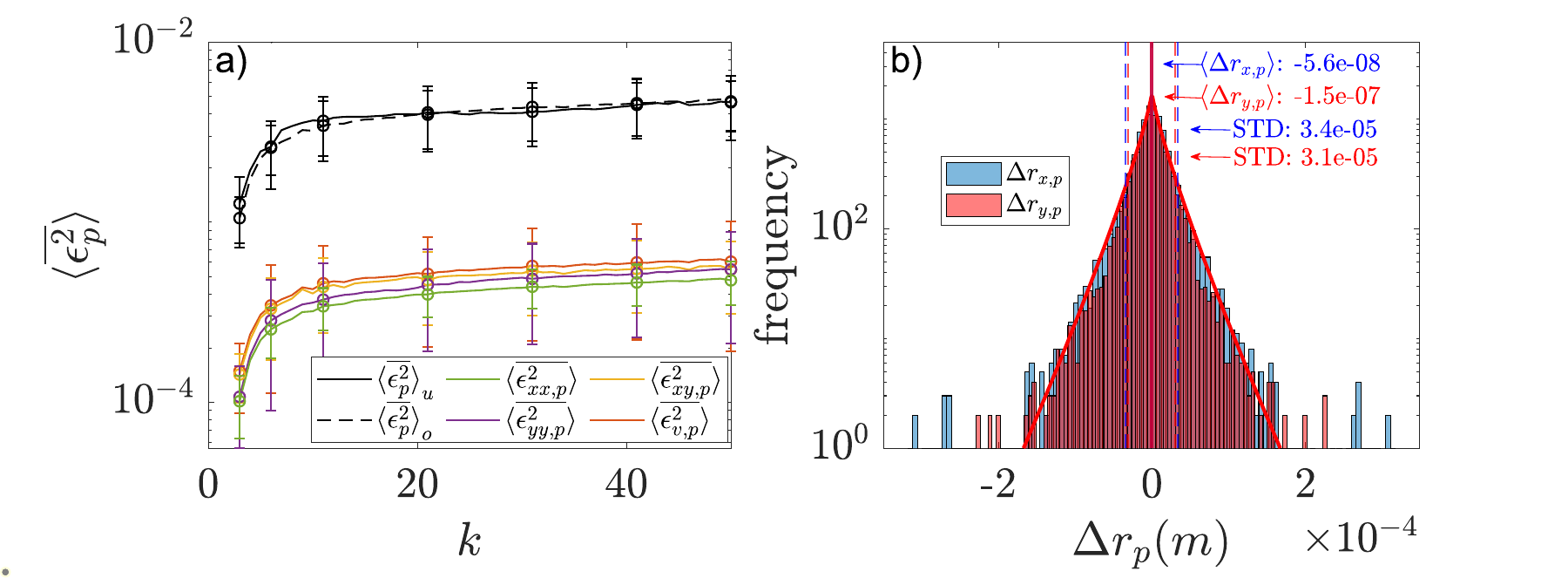}
    \caption{a) \textcolor{black}{Cumulative} quadratic plastic strain components as a function of the number of oscillations. b) Distribution of the relative displacements $\langle \Delta r_{x} \rangle$ and $\langle \Delta r_{y} \rangle$ between cycles $k_r$ and $50$. In both panels a) and b), the data correspond to $\theta_m = 16.8^\circ$, position $P1$, and $k_r = 2$.}
    \label{fig:sup:quad}
\end{figure}

\subsection{Quadratic terms from Bagi cells and DAWS}
The \textcolor{black}{cumulative plastic} quadratic strain terms obtained from the two methods are plotted in Figure~\ref{fig:sup:quad} a) for the interval $[a_{k_r},a_{k}]$ and $\theta_m = 16.8^\circ$ at position $P1$. The quantities $\langle \overline{\epsilon_{xx,p}^2} \rangle$, $\langle \overline{\epsilon_{yy,p}^2} \rangle$, $\langle \overline{\epsilon_{xy,p}^2} \rangle$, and $\langle \overline{\epsilon_{v,p}^2} \rangle$ in Bagi cells similarly increase with the number of oscillations and are all of the same order of magnitude. Their sum, $\langle \overline{\epsilon^2}\rangle_{o} = \left(\langle \epsilon_v^2 \rangle + 2 \times \sum_{i,j} \langle \epsilon_{i,j}^2 \rangle \right)$ ($i,j \neq z$), can be compared to $\langle \overline{\epsilon^2} \rangle_{u}$, which is derived from the inversion of Eq.~\ref{bicout:eq2b} and includes all squared components of the strain tensor. Figure~\ref{fig:sup:quad} a) shows that $\langle \overline{\epsilon^2}\rangle_{o}$ and $\langle \overline{\epsilon^2} \rangle_{u}$ are in good agreement. Therefore, comparing our experimental results with the theoretical predictions from~\cite{bicout1994} (Eq.~\ref{bicout:eq2b}), suggests that the quadratic terms $\langle \overline{\epsilon_{xz}^2} \rangle$, $\langle \overline{\epsilon_{yz}^2} \rangle$, and $\langle \overline{\epsilon_{zz}^2} \rangle$ can be neglected.

A comparison between both methods is also possible via the estimate of the root-mean-square relative displacements between grains. From the imaging method, computing the relative displacement $\Delta r$ between nearest neighbors yields the distributions $\Delta r_{x}$ and $\Delta r_{y}$, as shown in Figure~\ref{fig:sup:quad}b). \textcolor{black}{The distributions are stretched exponentials $P(x) = A \exp\left[-\left(|x|/B\right)^\beta\right]$. The fitted parameters are $A=3.1\times 10^4$, $B=6.5\times 10^{-5}$ and $\beta=0.86$, generally indicating deviations from affinity due to geometrical exclusion~\cite{radjai2002}.} We find $\langle \overline{\Delta r_x^2} \rangle_o^{1/2} \approx 34~\mu$m and $\langle \overline{\Delta r_y^2} \rangle_o^{1/2} \approx 31~\mu$m. Turning to DAWS, we calculate from Eq.~\ref{bicout:eq2a} and~\ref{bicout:eq2b} that in all cases $V(\Delta \Phi) \gg \langle \Delta \Phi \rangle^2$. Hence, $V(\Delta \Phi) \approx \langle \Delta \Phi^2 \rangle \sim q^2 \langle \Delta r^2 \rangle_u / 3$~\cite{weitzpage}. From Figure~\ref{fig:sup:quad}a), we find that $\langle \overline{\epsilon^2} \rangle_u \sim 10^{-3}$, which leads to $\langle \overline{\Delta r^2} \rangle_u^{1/2} \sim 47~\mu$m, similar to the value obtained from the optical method. It is important to note that the relative mean squared displacement $\langle \Delta r^2 \rangle_u$ measured by multiply scattered waves does not depend on the symmetric strain tensor and is therefore distinct from the more commonly used mean square displacement derived from the displacement gradient tensor.

In summary, DAWS and the optical method based on Bagi cells provide consistent independent measurements of strain fluctuations. The optical method has the advantage of allowing the calculation of average deformation but is limited to the study of grains in contact with the solid walls. In contrast, DAWS captures the sum of all squared strain terms in the bulk but cannot measure the average strain.

\subsection{Strain fluctuations : Intracycle \textcolor{black}{variability} and plasticity.}
\begin{figure}[htbp]
    \centering
    \includegraphics[scale=0.57]{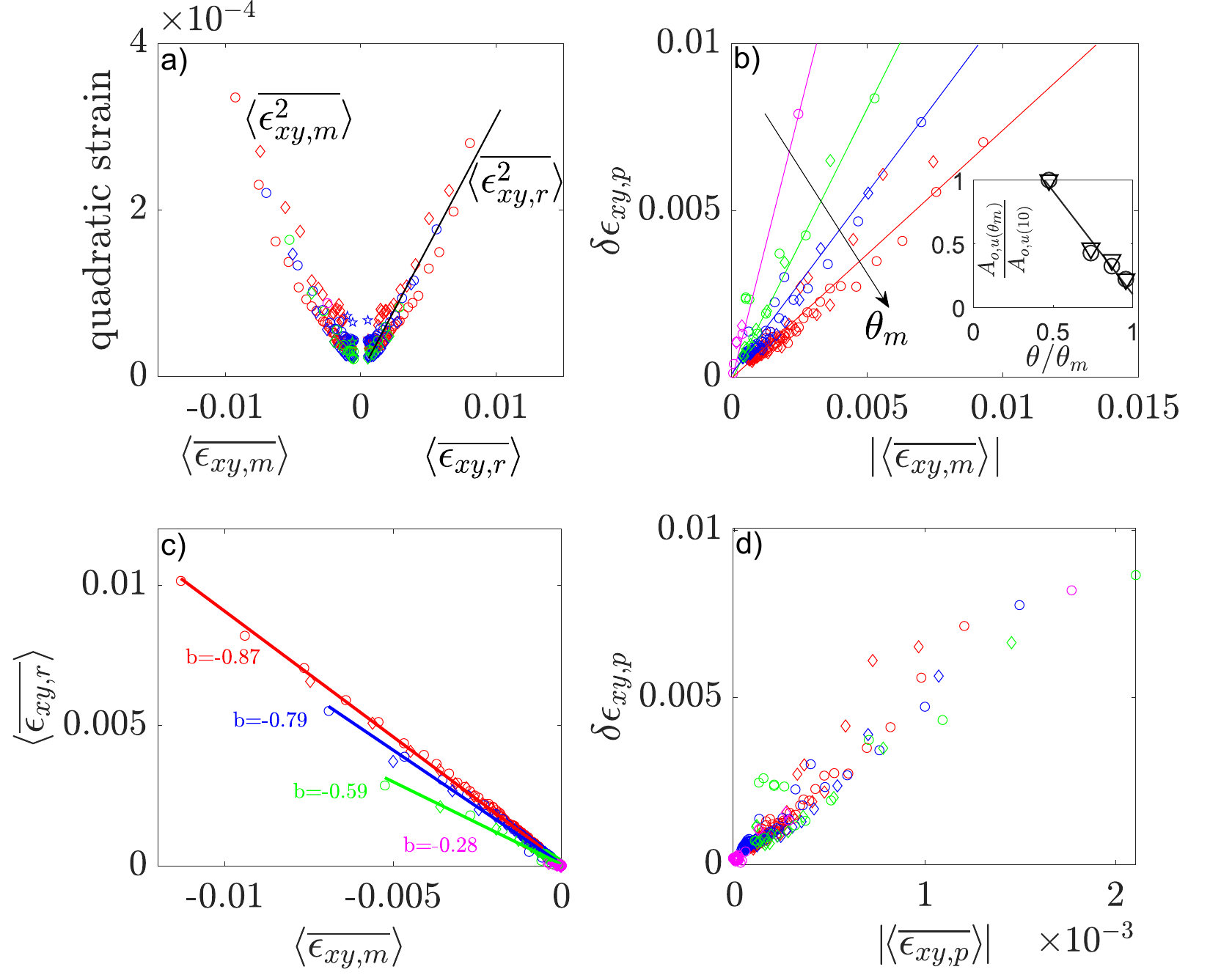} 
    \caption{a) Mean square strain from Bagi cells as a function of the average shear strain $\langle \overline{\epsilon_{xy,m}} \rangle$ and the average backward shear strain $\langle \overline{\epsilon_{xy,r}} \rangle$. b) Plastic strain fluctuations from Bagi cells as a function of the maximum strain $|\langle \overline{\epsilon_{xy,m}} \rangle|$. Inset of b): Circles represent $A_o(\theta_m)$ from Eq.~\ref{azero} (Bagi cells), normalized by $A_o(10.8^\circ) = 3.3$. Triangles represent $A_u(\theta_m) / A_u(10.8^\circ)$ from DAWS (see Figure~\ref{fig:plastic-rescale2}b in Appendix~\ref{sec:daws}), with $A_u(10.8^\circ) = 8.3$. The line is a linear fit to the data: $A_{o,u}(\theta_m) / A_{o,u}(10.8^\circ) = 1.6 \times \left(1 - \frac{\theta_m}{\theta_r}\right)$. \textcolor{black}{c) Mean strain for the backward branch of the shear cycles $\langle \epsilon_{xy,r}\rangle$ as a function of the forward 
    branch $\langle \epsilon_{xy,m}\rangle$. \textcolor{black}{d) Fluctuations of plastic strain as a function of the mean strain.}}
    Circles correspond to position $P1$, and diamonds to position $P3$. Position $P2$, represented by stars, is plotted only in panel a). The colors red, blue, green, and violet correspond to measurements at $\theta_m = 21.8^\circ$, $19.8^\circ$, $16.8^\circ$, and $10.8^\circ$, respectively. 
    }
        \label{fig:plastic-rescale}
\end{figure}
Since all measured quadratic strain terms are equivalent, and for the sake of conciseness, the remainder of the paper focuses on describing one specific strain component: the shear component (optical method). In particular, we \textcolor{black}{will show} how the \textcolor{black}{intracycle plastic} shear strain fluctuations \textcolor{black}{$\delta\epsilon_{xy,p}$} between $[a_k, a_{k+1}]$ \textcolor{black}{emerge} from the consecutive forward (maximum) strain $\epsilon_{xy,m}$ over the interval $[a_k, c _k]$ and the backward (return) strain $\epsilon_{xy,r}$ over $[c_{k}, a_{k+1}]$ (see sketch) in Fig.~\ref{fig:cycle}b). \textcolor{black}{Data from the optical method are presented in the main text, while complementary data from DAWS, leading to the same conclusions and suggesting that the observed effects are not confined to the possible peculiarities of grains at the solid walls, are reported in Appendix~\ref{supp:quad}.}

The variation of the quadratic strain as a function of both $\langle \overline{\epsilon_{xy,m}}\rangle$ and $\langle \overline{\epsilon_{xy,r}}\rangle$ from Bagi cells is plotted in Figure~\ref{fig:plastic-rescale}a) for all three zones of interest and the four oscillation angles. \textcolor{black}{Although each mean strain corresponds to different states of compaction and applied stress, all data collapse onto a single curve, approximated by $\langle \overline{\epsilon_{xy,r,m}^2}\rangle = D_o \times |\langle \overline{\epsilon_{xy,r,m}}\rangle|^\gamma$, with $D_o \approx 0.025$ an effective diffusion coefficient and \(\gamma \simeq 1\)}, consistent with random motion driven by shear~\cite{utterdiff, mueth2003, robbins3}. We note, however, that  data spanning several orders of magnitude in strain amplitude would be necessary to determine a more reliable estimate of the $\gamma$ exponent. Importantly, the term "diffusion" is used here to describe the proportionality between strain fluctuations and mean strain. 

Following the completion of each forward–backward shear cycle, the fluctuations of the intracycle plastic strain can be monitored across successive cycles, as shown in Fig.~\ref{fig:plastic-rescale}b, as a function of the maximum shear strain. \textcolor{black}{Data from position P$2$ are not discussed because the values are too close to the noise level $\sim 10^{-4}$.} First, for a fixed oscillation angle $\theta_m$, the fluctuations grow linearly with the mean maximum strain. Second, at a fixed maximum strain, larger oscillation angles result in smaller fluctuations. This can be described by
\begin{equation}
\label{azero}
\delta \epsilon_{xy,p}\approx\langle \overline{\epsilon_{xy,p}^2} \rangle^{1/2} \sim A_o |\langle \overline{\epsilon_{xy,m}} \rangle|
\end{equation}
$A_o$ is a function of $\theta_m$ and is plotted in Figure~\ref{fig:plastic-rescale}~b) (inset), corresponding to a linear decrease $A_{o}(\theta_m) \propto \left(1-\frac{\theta_m}{\theta_r}\right)$. As shown in Figure~\ref{fig:plastic-rescale2} in Appendix~\ref{quad_app}, we observe similar trends with DAWS. The parameter $A_u$, which is the DAWS equivalent of $A_o$, exhibits the same behavior as $A_o$ (Figure~\ref{fig:plastic-rescale}~b), inset). 

\begin{figure}[h]
    \centering
    \includegraphics[scale=0.6]{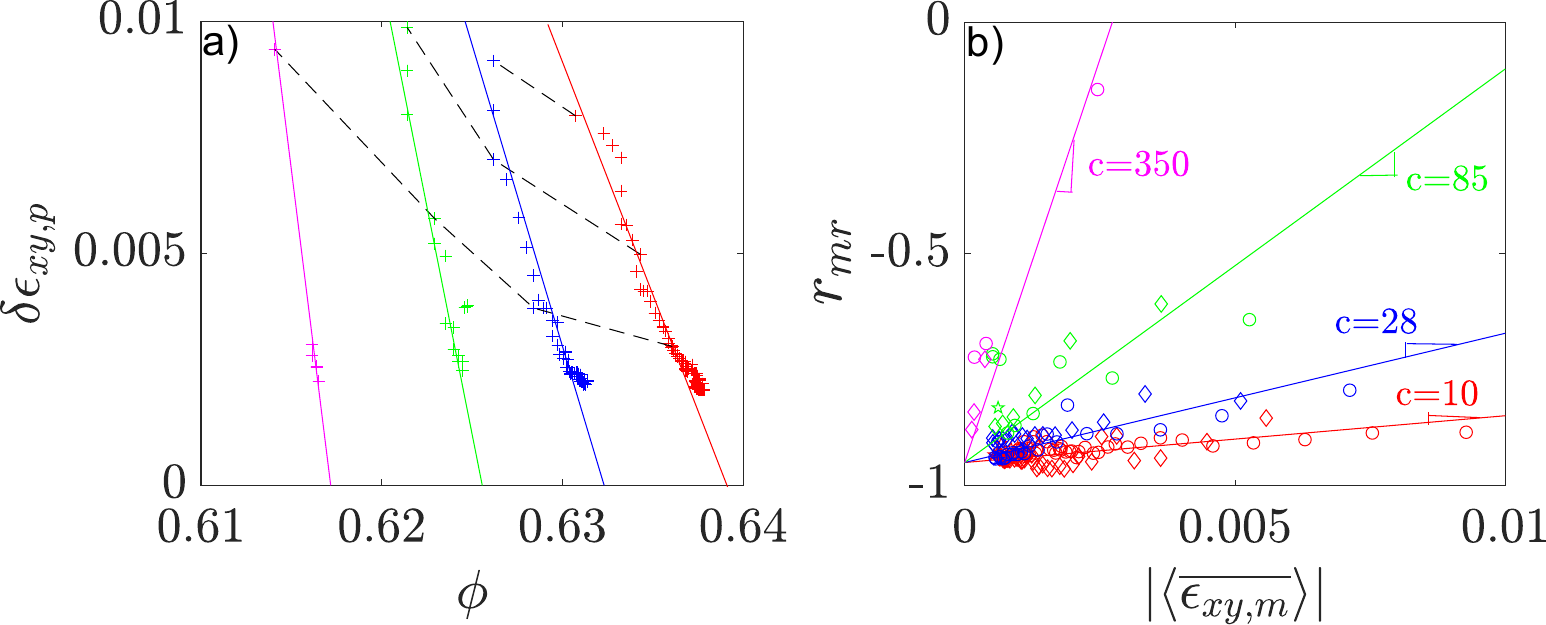}
    \caption{a) Fluctuations of plastic strain as a function of the density of the packing estimated from the compaction near the free surface (see text). The crosses are the experimental points smoothed with a moving average of 4 points. Dashed lines represent iso-strain points, from bottom to top $\langle \overline{\epsilon_{xy,m}}\rangle=2.5\times 10^{-3},5.2\times 10^{-3},6.9\times 10^{-3}$. Plain lines are guides to the eye. b) Correlation between $\epsilon_{xy,m}$ and $\epsilon_{xy,r}$ strains in a shear cycle, with the slope $c$ indicated next to the linear fit of the data for each angle. Red, blue, green, and pink lines and points correspond to $\theta_m=21.8^\circ, 19.8^\circ, 16.8^\circ$ and $10.8^\circ$, respectively.}
        \label{fig:toy}
\end{figure}

With the constitutive relation $\epsilon_{xy}(\theta, \phi)$ (increasing function of $\theta_m$ and a decreasing function of $\phi$), Eq.~\ref{fig:fluctuations} reads $
\delta \epsilon_{xy,p} \sim A_o(\theta_m) \times |\langle \overline{\epsilon_{xy,m}} \rangle|(\theta_m,\phi(\theta_m,k))$. \textcolor{black}{It is therefore useful to clarify the combined effects of stress, strain and density in the emergence of plastic strain fluctuations}. 
At fixed $\theta_m$, \textcolor{black}{the packing fraction increases with $k$ so that $\epsilon_{xy}(\theta_m, \phi(\theta_m,k))$ decreases. From Eq.~\ref{azero} fluctuations of plastic strain decrease as a simple effect since no plasticity is expected at vanishing forward shear amplitude.} In order to examine the iso-density and iso-strain cases, $
\delta \epsilon_{xy,p}$ is re-plotted in Figure~\ref{fig:toy} \textcolor{black}{a)} as a function of $\phi$, estimated using the expression in~\ref{eq:compac} (the following analysis is not sensitive to small variations of $\phi_0$ and $h$, nor to the particular method used to calculate the packing fraction). Comparison of iso-$\phi$ data at $\phi \sim 0.65$ in Figure~\ref{fig:toy} a) for $\theta_m = 19.8^\circ$ and $21.8^\circ$ suggests that fluctuations increase with higher stress, and the stress-strain constitutive relation dominates over the moderate effect of $\theta_m$ on $A_o$. Finally, at fixed $\epsilon_{xy}(\theta, \phi)$ (iso-strain dashed lines in Figure~\ref{fig:toy} a)), the plastic strain fluctuations increase with decreasing $\theta_m$ via $A_o$, as could be observed in Figure~\ref{fig:plastic-rescale} b). This in turn forces $\phi$ to decrease. Thus, the higher iso-strain plastic fluctuations are associated with both lower packing fraction and lower stress level $\theta_m$. 

Figure~\ref{fig:plastic-rescale}~c) shows that $\theta_m$ is key. The forward and backward mean strains are related by a linear relationship independent of the packing fraction $\langle \overline{\epsilon_{xy,r}} \rangle = -b \langle \overline{\epsilon_{xy,m}} \rangle$ where $b = b(\theta_m)$ only, and $0 \leq b \leq 1$. Parameter $b \to 1$ (or $b \to 0$) corresponds to a more elastic (or plastic) behavior. As determined in Figure~\ref{fig:plastic-rescale}~c), \textcolor{black}{$b\approx 0.28, 0.59, 0.79$ and $0.87$} for $\theta_m=10.8^\circ, 16.8^\circ, 19.8^\circ~ \text{and}~21.8^\circ$, respectively, showing that larger oscillation angles increase the reversibility of the mean strain. \textcolor{black}{This behavior leads to Figure~\ref{fig:plastic-rescale}~d) where the variation of $\delta \epsilon_{xy,p}$ on its mean is established by rescaling the abscissa with $\langle \overline{\epsilon_{xy,p}} \rangle=(1-b)\langle \overline{\epsilon_{xy,m}} \rangle$. Remarkably, all data collapse onto a single master curve, revealing a unique dependence of the plastic fluctuations on the mean such that $\delta \epsilon_{xy,p} \sim C_o \left|\langle \overline{\epsilon_{xy,p}} \rangle\right|$, where $C_o \approx 5$ (see also Figure~\ref{fig:plastic-rescale2}c) for the equivalent DAWS measurements).}
\section{Discussion}
\label{sec:discussion}
We now establish the link between the results presented in Figure~\ref{fig:plastic-rescale}.  The plastic strain is $\epsilon_p = \epsilon_m + \epsilon_r$ so that treating the strain in the Bagi cell as a random variable constrains the variance to satisfy $V(\epsilon_p) = V(\epsilon_m) + V(\epsilon_r) + 2 \text{cov}(\epsilon_m, \epsilon_r)$, where $V$ represents the variance and $\text{cov}$ is the covariance. Along with $
\langle \overline{\epsilon_{xy,m,r}^2} \rangle \sim D_o |\langle \overline{\epsilon_{xy,m,r}} \rangle|$, this provides $
\langle \overline{\epsilon_{xy,p}^2} \rangle \approx \alpha |\langle \overline{\epsilon_{xy,m}} \rangle|$ where $\alpha = D_o (1 + b + 2b^{1/2} r_{mr})$ and $r_{mr}$ is the correlation coefficient, defined as the covariance divided by the product of the standard deviations of the forward and backward strains, $\epsilon_{xy,m}$ and $\epsilon_{xy,r}$. $r_{mr}$ can also be determined experimentally as in Figure~\ref{fig:toy} b). It is negative (indicating anticorrelation between maximum and backward shear) and varies linearly with the strain $r_{mr}=c|\langle \overline{\epsilon_{xy,m}}\rangle|-1$. The slope $c=c(\theta_m)$, annotated in Figure~\ref{fig:toy} a), is a decreasing function. Arranging the various terms above gives $\alpha=D_o\left(1+b-2b^{1/2}+2b^{1/2}c\times|\langle \overline{\epsilon_{xy,m}}\rangle|\right)$. 

With this expression for $\alpha$, fluctuations vanish as expected for $\epsilon \rightarrow 0$. However, for larger strains ($> 10^{-3}$), $\delta \epsilon_{xy,p} \sim b^{1/4} \left(2c D_o\right)^{1/2} |\langle \overline{\epsilon_{xy,m}} \rangle|$, which directly yields the expression for $A_o=b^{1/4} \left(2c D_o\right)^{1/2}$ by identification with Eq.~\ref{azero}. In our experiments, $D_o$ is constant, and $b^{1/4}$ varies slowly, so $A_o$ is primarily controlled by the parameter $c$. As a consistency check, $A_o$ is plotted as a function of $c^{1/2}$ in Figure~\ref{fig:13} in the Appendix~\ref{supp:param}. The linear fit $A_o = 0.17 \times c^{1/2}$ is in good agreement with the independent estimate $b^{1/4} \left(2 D_o\right)^{1/2} \approx 0.2$. \textcolor{black}{The relation between the variance and the mean of the plastic strain eventually reads $\delta \epsilon_{xy,p} \sim C_o \left|\langle \overline{\epsilon_{xy,p}} \rangle\right|$, where $C_o = b^{1/4} (2 D_o c)^{1/2} \left(1 - b\right)^{-1}$ remains close to $5$. This enforces $c\propto\frac{(1-b)^2}{b^{1/2}}$ to be a monotonically decreasing function of $b$.}

\textcolor{black}{We now outline a speculative scenario to explore the functional relationship between $c$ and $b$. As an analogy with a granular medium, we consider a system with a finite number of mesoscopic “sites” that accumulate strain under forward shear $\epsilon_m$. Here, $D_0$ represents the strain scale of independent deformation sites, matching the variability of forward and backward strains with their mean and the order of magnitude of the largest strain cells shown in Figure~\ref{fig:fluctuations}. During reverse shear, the sites exhibit an average elastic recovery $-b\epsilon_m$ to which a stochastic term proportional to $+c\epsilon_m$ is added. $b(\theta_m)$ quantifies the effective proportion of reversible sites and hence represent a memory parameter while $c(\theta_{m})$ controls how increasing shear amplitude produces random irreversible deformation. Then, the creation of reversible sites through shearing at higher load exhausts the reservoir of sites that can still slip irreversibly. More work is needed to investigate the potential link of our results with the emergence of limit cycles~\cite{keim2019}.}

Plasticity \textcolor{black}{in our system, viewed through the lens of local fluctuations, could be compared to other divided systems}. In emulsions~\cite{munch97}, fluctuations increase with decreasing packing fraction at controlled strain. In colloidal glasses~\cite{petekidis2002}, low packing fraction leads to fewer plastic events, as the lubricated interactions between grains allow for reversibility. Our study suggests distinct signatures of emerging plasticity, shedding light on the effects of applied load and memory.
\section{Conclusion}
We have experimentally studied, using a setup combining optical analysis and DAWS, the fluctuations of strain in a granular medium subjected to oscillatory stress. A detailed analysis of the sources of variability suggests that the effect of the gradient of the mean strain  along the vertical is much smaller than the strain fluctuations at the scale of the grain $d$. \textcolor{black}{This allows to compare both optical and DAWS methods which, providing similar results, suggest that the observations made by measuring the strain fluctuations through imaging the grains at the wall are relevant in the bulk.}
We find that the variance of the local strain is proportional to its mean, consistent with observations in discrete or partitioned systems where mutual exclusion is a key governing mechanism. Furthermore, the fluctuations of plastic strain depend solely on the mean, regardless of the compaction state or the applied load, and arise from the interplay between a memory term and strain anticorrelations. \textcolor{black}{Quantifying scaling behaviors~\cite{radjai2002,chechkin2017} across broader strain ranges and multiple system sizes, as well as investigating the effect of polydispersity~\cite{jiang2023} on strain variability, represent relevant directions for future research.}
\subsection*{Acknowledgements}
We thank D. Hautemayou, C. Mézière and P. Moucheront for their help in building the experimental setup. This work was funded by Labex MMCD.  
\subsection*{Author contribution statement}
All authors contributed equally to the paper.
\subsection*{Data availability statement}
The data that support the findings of this study can be obtained from the corresponding author upon reasonable request.
\nocite{}
\bibliographystyle{unsrt}

\begin{thebibliography}{10}
	
	\bibitem{argon2}
	A.S. Argon and H.Y. Kuo.
	\newblock {\em Materialss Science and Engineering}, 39:101, 1979.
	
	\bibitem{chen}
	D.~Chen, K.W. Desmond, and E.R. Weeks.
	\newblock {\em Soft Matter}, 8:10486, 2012.
	
	\bibitem{spaepenW}
	K.E. Jensen, D.A.Weitz, and F.~Spaepen.
	\newblock {\em Phys. Rev. E.}, 90:042305, 2014.
	
	\bibitem{combe2015}
	G.~Combe, V.~Richefeu, M.~Stasiak, and A.~Atman.
	\newblock {\em Phys. Rev. Lett.}, 115:238301, 2015.
	
	\bibitem{radjai2002}
	F.~Radjai and S.~Roux.
	\newblock {\em Phys. Rev. Lett.}, 89:064302, 2002.
	
	\bibitem{cao2018}
	Y.~Cao et~al.
	\newblock {\em Nat. Commun.}, 9:2911, 2018.
	
	\bibitem{pouliquen2003}
	O.~Pouliquen, M.~Belzons, and M.~Nicolas.
	\newblock {\em Phys. Rev. Lett.}, 91:014301, 2003.
	
	\bibitem{hurley}
	C.~Zhai, N.~Albayrak, J.~Engvist, S.~Hall, J.~Wright, M.~Majkut, E.B. Herbolt,
	and R.C. Hurley.
	\newblock {\em Phys. Rev. E}, 105:014904, 2022.
	
	\bibitem{hohler1997}
	R.~{H\"ohler}, S.~Cohen-Addad, and H.~Hoballah.
	\newblock {\em Phys. Rev. Lett.}, 79:1154, 1997.
	
	\bibitem{munch97}
	P.~H\'ebraud, F.~Lequeux, J.P. Munch, and D.J. Pine.
	\newblock {\em Phys. Rev. Lett.}, 78:4657, 1997.
	
	\bibitem{petekidis2002}
	G.~Petekidis, A.~{Moussa\"id}, and A.N. Pusey.
	\newblock {\em Phys. Rev. E}, 66:051402, 2002.
	
	\bibitem{crassous2}
	A.~Amon, V.~Bau Nguyen, A.~Bruand, J.~Crassous, and E.~Clément.
	\newblock {\em Phys. Rev. Lett.}, 108:135502, 2012.
	
	\bibitem{crassous1}
	A.~Le Bouil, A.~Amon, S.~McNamara, and J.~Crassous.
	\newblock {\em Phys. Rev. Lett.}, 112:246001, 2014.
	
	\bibitem{durian_97}
	N.~Menon, , and D.J. Durian.
	\newblock {\em Science}, 275:1920, 1997.
	
	\bibitem{weitzpage}
	M.L. Cowan, I.P. Jones, J.H. Page, and D.~A. Weitz.
	\newblock {\em Phys. Rev. E}, 65:066605, 2002.
	
	\bibitem{leosoft2020}
	J.~L\'eopold\`es and X.~Jia.
	\newblock {\em Soft Matt.}, 16:10716, 2020.
	
	\bibitem{culling}
	W.E.H. Culling.
	\newblock {\em J. Geol.}, 71:127, 1963.
	
	\bibitem{choi}
	J.~Choi, A.~Kudrolli, R.R. Rosales, and M.Z. Bazant.
	\newblock {\em Phys. Rev. Lett.}, 92:174301, 2004.
	
	\bibitem{repose}
	H.M. Jaeger, C.~Liu, and S.R. Nagel.
	\newblock {\em Phys. Rev. Lett}, 62:266001, 1989.
	
	\bibitem{stardist}
	Uwe Schmidt, Martin Weigert, Coleman Broaddus, and Gene Myers.
	\newblock Cell detection with star-convex polygons.
	\newblock {\em CoRR}, abs/1806.03535, 2018.
	
	\bibitem{track}
	J.C. Crocker and D.G. Grier.
	\newblock {\em J. Colloid Interface Sci.}, 179:298, 1996.
	
	\bibitem{bagi96}
	K.~Bagi.
	\newblock {\em Mechanics of Materials}, 22:165, 1996.
	
	\bibitem{savage93}
	S.B. Savage and R.~Dai.
	\newblock {\em Mechanics of materials}, 16:225, 1993.
	
	\bibitem{sheng}
	P.~Sheng.
	\newblock {\em Introduction to Wave Scattering, Localization and Mesoscopic
		Phenomena}.
	\newblock Springer, 2006.
	
	\bibitem{bicout1994}
	D.~Bicout and R.~Maynard.
	\newblock {\em Physica A}, 199:387, 1993.
	
	\bibitem{nagel2}
	E.R. Nowak, J.B. Knight, E.~Ben-Naim, H.M. Jaeger, and S.~R. Nagel.
	\newblock {\em Phys. Rev. E.}, 57:1971, 1998.
	
	\bibitem{hicher}
	Bernard Cambou and Pierre-Yves Hicher.
	\newblock {\em Elastoplastic Modeling of Soils: Cyclic Loading}, chapter~4,
	pages 143--186.
	\newblock John Wiley and Sons, Ltd, 2008.
	
	\bibitem{utterdiff}
	B.~Utter and R.P. Berhinger.
	\newblock {\em Phys. Rev. E}, 69:031308, 2004.
	
	\bibitem{mueth2003}
	D.M. Mueth.
	\newblock {\em Phys. Rev. E.}, 67:011304, 2003.
	
	\bibitem{robbins3}
	C.E. Maloney and M.O. Robbins.
	\newblock {\em J. Phys. Condens. Matter}, 20:244128, 2008.
	
	\bibitem{keim2019}
	N.~M. Keim, J.D. Paulsen, Z.~Zeravcic, S.~Sastry, and S.D. Nagel.
	\newblock {\em Rev. Mod. Phys.}, 91:035002, 2019.
	
	\bibitem{chechkin2017}
	A.~V. Chechkin, F.~Seno, R.~Metlzer, and I.~Sokolov.
	\newblock {\em Phys. Rev. X}, 7:021002, 2017.
	
	\bibitem{jiang2023}
	D.~M.~Sussman Y.~Jiang and E.~R. Weeks.
	\newblock {\em Phys. Rev. E}, 108:054605, 2023.
	
	\bibitem{kyto}
	H.K. Kyt{\"{o}}maa.
	\newblock {\em Powder Technology}, 82:115, 1995.
	
\end{thebibliography}

\clearpage
\appendix
\section{DAWS}
\label{sec:daws}
\begin{figure}
	\centering
	\includegraphics[scale=0.5]{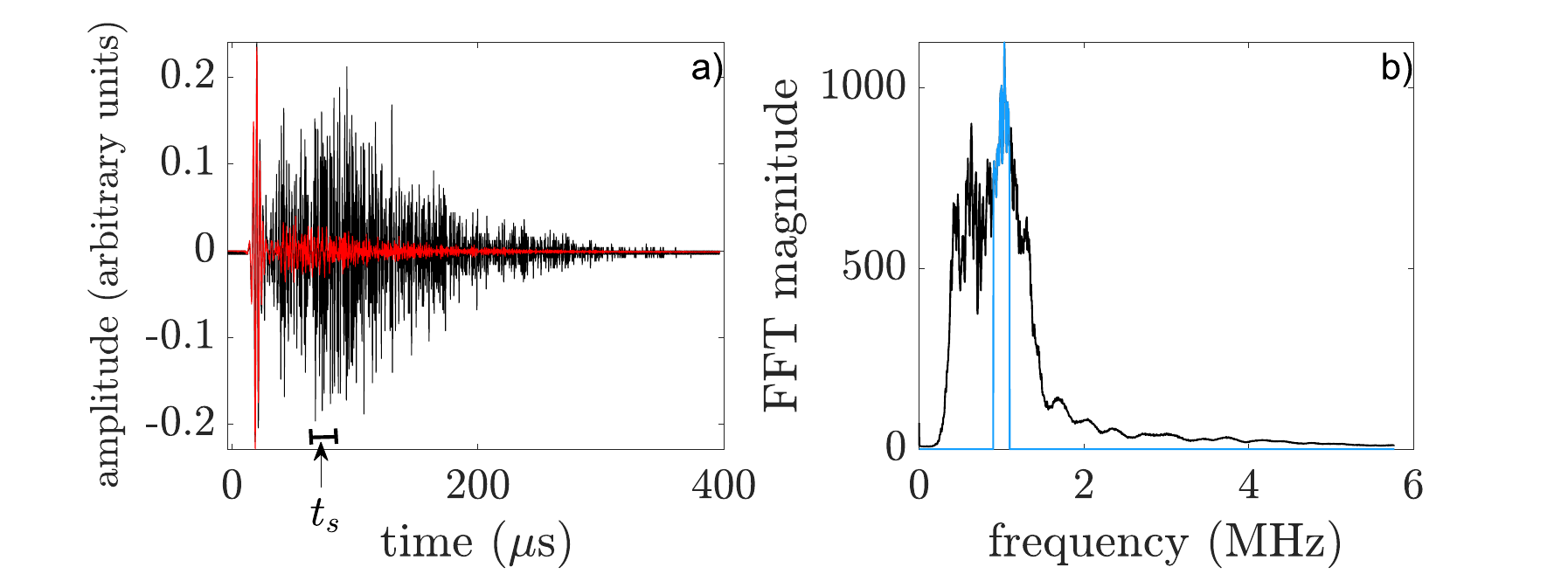}
	\caption{(a) The black curve represents a typical signal transmitted through the granular suspension, while the red curve corresponds to the coherent part, obtained by averaging $10^2$ signals from different packings after multiple full rotations of the drum. The window used to compute the signal decorrelation is taken at $t_s$ (see text).  
		(b) Average spectrum of the transmitted signal, with the filtering window (in cyan) centered at $1$ MHz $\pm 10\%$.}  
		\label{fig:exsignal}
\end{figure}

With an appropriate rotation of the drum, the emission-reception transducers can be positioned such that their axis passes through the center of the study areas ($P1$, $P2$, or $P3$ in Figure~\ref{fig:image}~a)). The recorded signals consist of a coherent component, while the scattering effects are captured in the late "coda" part, as illustrated in Figure~\ref{fig:exsignal}~a).  
The coherent part (red signal in Figure~\ref{fig:exsignal}~a)) is obtained by averaging one hundred signals transmitted through independent configurations, each acquired after several full rotations of the drum. This coherent wave is subtracted from raw signals, which are subsequently filtered to $1$ MHz~$\pm 10\%$ (see Figure~\ref{fig:exsignal}~b)).  

When the drum oscillates below the angle of repose, the relative motion of the grains induces changes in the ultrasound signals. The coherent part of the signal depends on the average properties of the medium, such as the effective bulk modulus and density.   
For the calculation of correlations between signals, we select a window between $65~\mu$s and $85~\mu$s, centered around the propagation time $t_s$, which corresponds to the maximum of the transmitted intensity (see Figure~\ref{fig:exsignal}~a)), as obtained from the envelope of $10^2$ signals from different configurations. This time can be interpreted as the time required for scattered waves to probe a volume of the order of $L^3$, where $L$ is the thickness of the drum. Indeed, scattered waves explore a typical length scale diffusively, given by $\sqrt{6\mathcal{D}t}$, where $\mathcal{D}$ is the diffusion coefficient of the waves. With an order of magnitude estimate $\mathcal{D} \sim 1$~mm$^2/\mu$s~\cite{weitzpage} and $L=20$~mm, we obtain a typical timescale $t_D\sim 70$~$\mu$s, comparable to $t_s$. The number of scattering events along the acoustic path is estimated as $n \sim t_s v_e / l$, where $t_s$ is the time associated with the chosen temporal window, and $v_e$ is the energy velocity. Assuming $v_e \approx v$, the sound speed in water, and $\lambda \approx l$, $n \sim t_s f=80$ ($f$ is the frequency).   

Unlike previous studies in the incompressible limit~\cite{weitzpage}, the oscillations of the drum in our experiment induce compaction in the granular medium. This compaction has a small yet significant effect on the $\cos$ term in Eq.~\ref{bicout:eq}, requiring a correction of approximately $10\%$ to $g_1$ for accurate calculation of the phase variance. Using the model proposed by~\cite{kyto}, we have verified that the volumetric strain estimated from variations in the time of flight (coherent component of the acoustic signals) is consistent with values obtained through grain tracking (Bagi cells). The correction to $g_1$ based on volumetric strain can be applied using either of these two measurements.

\newpage
\section{Fitting parameters for the global compaction}
\begin{figure}[h]
    \centering
    \includegraphics[scale=0.5]{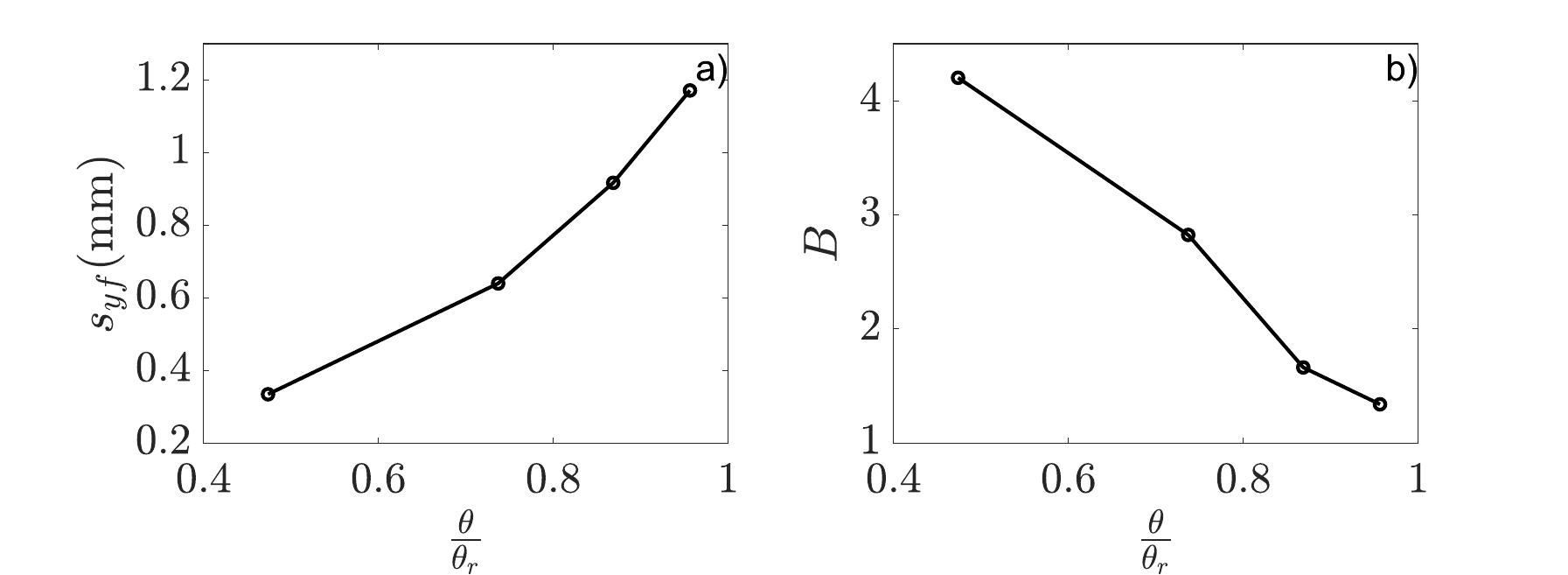}
    \caption{Parameters obtained from fitting, with Equation~\ref{eq:fit:nag}, the displacement of the free surface during compaction at different oscillation angles $\theta_m$. The oscillation angle is normalized by the angle of repose $\theta_r$.}
    \label{fig:sup:fitmodelnagel}
\end{figure}
\label{supp:paramFit}

For the material located between the free surface and of height $h$ the global packing fraction \(\phi\), which represents a measure of the compaction fraction, is related to the volumetric strain \(\epsilon_v\), by :
\[
\epsilon_v = \frac{\Delta V}{V_0} = -\frac{\Delta \phi}{\phi_0} \sim -\frac{s_y}{h},
\]
with $V_0$, $\phi_0$ the volume and the packing density of the reference packing. $s_y$ is the displacement of the free surface. This amounts to
\[
\phi = \frac{\phi_0 (s_y + h)}{h}.
\]

With the expression in~\cite{nagel2} :
\[
\phi = \phi_f - \frac{\phi_f-\phi_0}{1 + B \log\left(1 + \frac{t}{\tau}\right)},
\]

we obtain :

\begin{equation}
s_y = s_{yf} - \frac{s_{yf}}{1 + B \log(1 + k)},
\label{eq:fit:nag}
\end{equation}

This equation was used to fit our experimental data describing the displacement of the free surface during the compaction oscillations. The variable \(s_y\) represents the position of the surface during the compaction cycle \(k\), while \(s_{yf}\) corresponds to the final position of the surface after 50 cycles of compaction, reflecting the most compact state reached during the process.  

Higher values of \(s_{yf}\) indicate more densely compacted systems at the end of the compaction cycles, as inferred from the final position of the free surface. The dimensionless parameter \(B\) characterizes the kinetics of compaction, where lower values of \(B\) correspond to faster compaction dynamics.
\clearpage
\section{Quadratic strain from DAWS}
\label{quad_app}
\begin{figure}[htbp]
    \centering
    \includegraphics[scale=0.45]{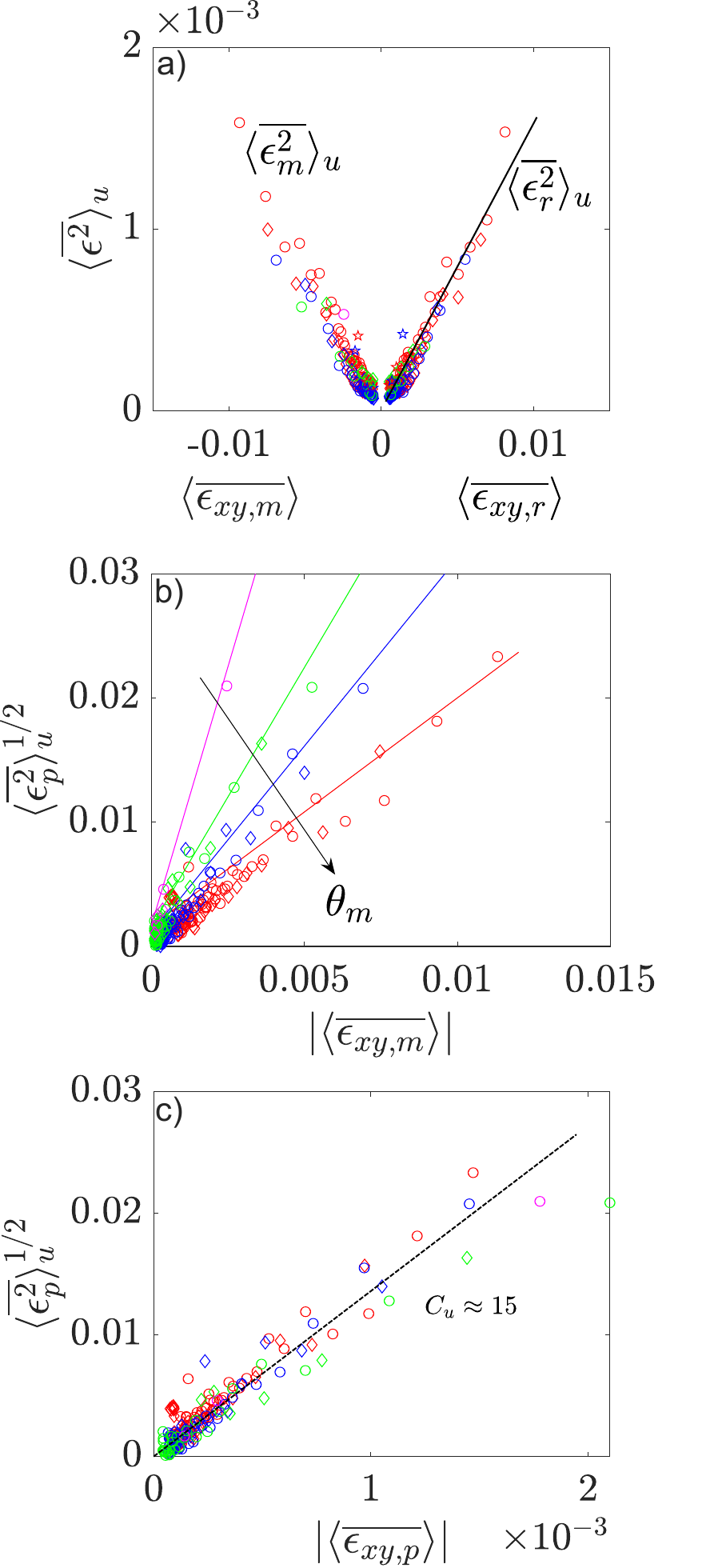} 
    \caption{a) Mean square strain from DAWS plotted versus the average shear strain $\langle \overline{ \epsilon_{xy,m}}\rangle$ and average backward shear strain $\langle \overline{\epsilon_{xy,r}}\rangle$, corresponding to pairs $[a_k,c_{k}]$ and $[c_k,a_{k+1}]$, respectively. b) Fluctuations of plastic strain from DAWS as a function of maximum strain $|\langle \overline{ \epsilon_{xy,m}}\rangle|$. Circles: position $P1$. Losange: position $P3$. Stars: position $P2$. Red, blue, green and violet represent measurements at $\theta_m=21.8^\circ, 19.8^\circ, 16.8^\circ$ and $10.8^\circ$ respectively. c) Fluctuations of plastic strain from DAWS as a function of mean plastic strain. In b) and c), only positions $P1$ and $P3$ are represented. \textcolor{black}{For a), b), and c) the values of $\langle \overline{\epsilon_{xy,m}} \rangle$   ,   $\langle \overline{\epsilon_{xy,r}} \rangle$   and   $\langle \overline{\epsilon_{xy,p}} \rangle$   used for building the abscissa are the same as in Figure 7 a)  b) and d).}}
    \label{fig:plastic-rescale2}
\end{figure}

In Figure~\ref{fig:plastic-rescale2} a), the sums of the quadratic terms obtained from DAWS (three positions, four angles) are plotted as a function of the mean strain measured in Bagi cells in the same conditions. We obtain $\langle \overline{\epsilon^2}\rangle_u = D_u \times |\langle \overline{\epsilon_{xy,r,m}}\rangle|$ with $D_u \approx 0.17$. The ratio $D_o/D_u\sim 10^{-1}$ is as expected, confirming the agreement between the quadratic terms from the optical method and DAWS.

With DAWS, we also found a linear relation $\langle \overline{\epsilon_{p}^2} \rangle_u^{1/2} = A_u |\langle \overline{\epsilon_{xy,m}} \rangle|$ as shown in Figure~\ref{fig:plastic-rescale2} b). For better comparison, $A_u=A_u(\theta_m)$ is plotted together with $A_o$ in the inset of Figure~\ref{fig:plastic-rescale} b) (main text), and has a similar dependency on the oscillation angle. Both methods show that, at fixed strain, the fluctuations of plastic strain decrease linearly with $\theta_m$.
\label{supp:quad}
\clearpage
\section{Linearity between parameters $c$ and $A_o$}
\label{supp:param}
\begin{figure}[h]
    \centering
    \includegraphics[scale=0.6]{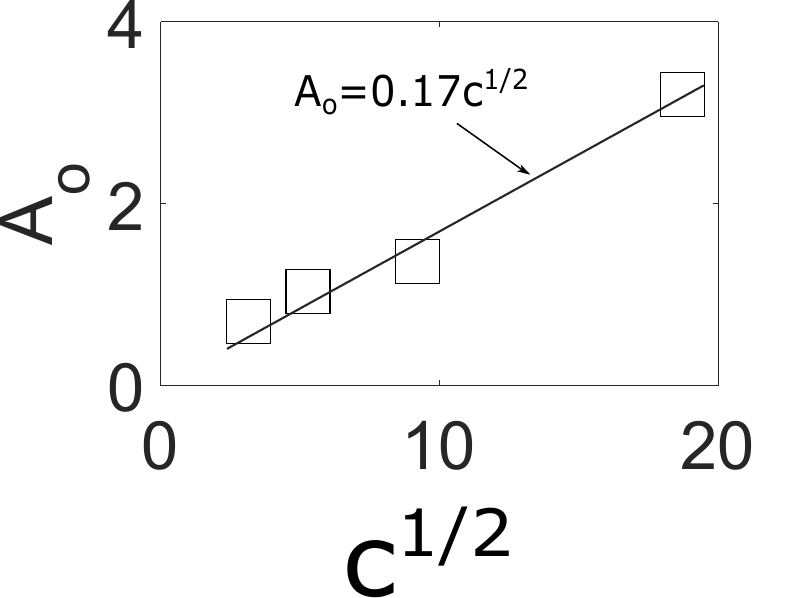}
    \caption{Parameter $A_o$ plotted versus $c^{1/2}$. Squares: experimental data. The plain line is a linear fit to the data.}
    \label{fig:13}
\end{figure}

\end{document}